\documentclass[a4paper]{article}

\usepackage[round]{natbib}
\usepackage[english]{babel}
\usepackage[utf8]{inputenc}
\usepackage[T1]{fontenc}

\usepackage{tcolorbox}

\usepackage{listings}
\usepackage{graphicx}
\usepackage[autostyle]{csquotes}

\usepackage[a4paper,top=3cm,bottom=2cm,left=3cm,right=3cm,marginparwidth=1.75cm]{geometry}

\usepackage{hyperref}
\usepackage{authblk}
\usepackage{rotating}
\usepackage{csvsimple}
\usepackage{float}

\usepackage{amsmath, blkarray}

\begin{document}


\title{\huge Knowledge-based Integration of Multi-Omic Datasets with Anansi:\\Annotation-based Analysis of Specific Interactions}
\author[1, 2, +]{Thomaz F. S. Bastiaanssen}
\author[3]{Thomas P. Quinn}
\author[1, 2]{John F. Cryan}

\affil[1]{\footnotesize APC Microbiome Ireland, University College Cork, Ireland.
}
\affil[2]{\footnotesize Department of Anatomy and Neuroscience, University College Cork, Ireland.
}
\affil[3]{\footnotesize Independent Scientist, Geelong, Australia
}
\affil[+]{\footnotesize Corresponding author: thomazbastiaanssen@gmail.com
}

\date{}

\Affilfont{\fontsize{4}{4}}

\maketitle

\begin{abstract} 
\textbf{Motivation:} Studies including more than one type of 'omics data sets are becoming more prevalent. Integrating these data sets can be a way to solidify findings and even to make new discoveries.  However, integrating multi-omics data sets is challenging. Typically, data sets are integrated by performing an all-vs-all correlation analysis, where each feature of the first data set is correlated to each feature of the second data set. However, all-vs-all association testing produces unstructured results that are hard to interpret, and involves potentially unnecessary hypothesis testing that reduces statistical power due to false discovery rate (FDR) adjustment. \\ \textbf{Implementation:} Here, we present the anansi framework, and accompanying R package, as a way to improve upon all-vs-all association analysis. We take a knowledge-based approach where external databases like KEGG are used to constrain the all-vs-all association hypothesis space, only considering pairwise associations that are \textit{a priori} known to occur. This produces structured results that are easier to interpret, and increases statistical power by skipping unnecessary hypothesis tests. In this paper, we present  the anansi framework and demonstrate its application to learn metabolite-function interactions in the context of host-microbe interactions. We further extend our framework beyond pairwise association testing to differential association testing, and show how anansi can be used to identify associations that differ in strength or degree based on sample covariates such as case/control status. \\
\textbf{Availability:}
\url{https://github.com/thomazbastiaanssen/anansi}

\end{abstract}

\newpage

\section{Introduction}

Techniques that aim to measure the totality of a certain type of biological molecules are known as 'omics. The most prevalent types of 'omics include (meta)genomics, (meta)transcriptomics, (meta)proteomics and metabolomics. 
In recent years, there has been an increase in studies that feature multiple types of 'omics data, which are referred to as multi-omics \citep{multi_omics}. For instance, in host-microbiome studies, it has become more common to measure both the microbial metagenome and the serum and/or gut metabolome. 'Omics approaches enable a broader and more exploratory avenue of doing research, potentially allowing the researcher to uncover complex patterns that would otherwise not have been discovered \citep{night_science}. However, dealing with big data sets comes with new challenges, especially with regard to the interpretation of results and the preservation of statistical power. 

Often, multi-omics analysis takes the form of pairwise all-vs-all association testing between the features of the data sets, an approach which inherits the same two challenges. First, an all-vs-all association procedure will produce unstructured results that are typically presented as a list of ``significant'' findings or a heatmap of associations. These lists and heatmaps can be difficult to interpret or generate new hypotheses from because the results are not put in the context of established biological knowledge. Second, the method can be wasteful in terms of statistical power. As every statistical test produces a p-value that ought to be adjusted (e.g., through false discovery rate (FDR) adjustment), if it is biologically unfeasible for an association to be real, testing for it anyway could be considered a waste of power and may result in false negatives. 

In this article, we present the anansi framework as an alternative approach to all-vs-all association testing that leverages knowledge databases to address the aforementioned challenges:
\begin{itemize}
    \item \textbf{Knowledge databases help structure results and improve interpretability by giving context to the results.} It is difficult to form hypotheses about results that are presented without context. 
    For instance, in the case of the microbiome, relating the levels of metabolites to the abundance of microbial species may result in \textit{significantly associated} metabolite-microbe pairs that cannot be explained from a biological perspective. 
    Rather than assessing microbes on a taxonomical level, one could instead assess the genes within those microbes that might encode for enzymes, receptors or other proteins that interact with those metabolites. By re-framing the analysis as a problem of protein-metabolite interactions we gain the ability to leverage our extensive knowledge of metabolic pathways when interpreting results and generating new hypotheses. 
    \item \textbf{Knowledge databases help improve statistical power by restricting the total number of tested hypotheses.} When performing an all-vs-all association analysis, many features that do not interact will still be tested for a statistically significant association. 
    Calculating additional p-values will result in a higher number of p-values to adjust by post-hoc methods like Benjamini-Hochberg's and Storey's q-value procedure, which will in turn lead to a loss of power \citep{BH,qvalue}. Conversely, assessing interactions in pairs of features that do not biologically interact risks encountering spuriously significant associations that do not lead to fruitful hypotheses.
\end{itemize}

Re-framing the point, if the goal of a multi-omics integration analysis is to identify real associations between the features of two biologically related data sets in order to formulate testable hypotheses, all-vs-all analysis may not always be the most appropriate approach. Here, we present the anansi (Annotation-based Analysis of Specific Interactions) framework and accompanying R package which uses knowledge databases to reduce unnecessary hypothesis testing, giving context to the results and improving statistical power. 
Although the method is general, we demonstrate its application on a microbiome-metabolme integration data set.

\newpage

\section{Related Work}

\subsection{Knowledge databases}
Anansi relies on the structure provided from knowledge databases. Typically, these are databases that that contain knowledge on features and how they interact, for example in the form of a molecular interaction network. Notable databases include \textbf{KEGG} \citep{kegg}, \textbf{MetaCyc} \citep{metacyc}, \textbf{CAZy} \citep{cazy2022}, \textbf{HMDB} \citep{hmdb} and \textbf{EggNOG} \citep{eggnog}. The main difference between these databases is the specific focus of their content and in many cases identifiers can be mapped from one database to another. 

\subsection{Measures of association}
Numerous methods exist to measure an association between two features. The most common are undoubtedly \textbf{Pearson's} and \textbf{Spearman's Rank} correlation coefficients. Both of these metrics can be thought of as special cases of a \textbf{linear model} \citep{kenney1962linear}. These methods are well-understood and perform well in general, though many types of 'omics data are compositional and inherently display negative correlations, for which these methods may yield spurious results \citep{gloor_frontiers}. Some methods have been introduced to specifically address these traits of compositional data, including \textbf{proportionality}, \citep{lovell2015proportionality, quinn2017propr}, a collection of metrics that investigate whether the ratio between two features remains stable. Other compositional methods include \textbf{SPARCC} \citep{SPARCC} and \textbf{SPieCeasi} \citep{SPieCeasi}, both of which are microbiome-oriented and assume a sparse matrix.  

\subsection{All-vs-all approaches}
The \textit{Hierarchical All-against-All association testing} (\textbf{HAllA}) framework aims to cluster features from the same data set in a data-driven manner before analysis in order to reduce the amount of testing, thus substantially improving power \citep{halla}. HAllA relies on data-driven clustering to reduce the amount of tests performed and thus the biological interpretation of these clusters 
is not guaranteed.
Notably, HAllA is designed for large population studies and confounding factors are thus expected to be \textit{regressed out} before analysis. 

\subsection{Ordination and learning-based approaches}
On a different axis, the \textbf{MINT} and \textbf{DIABLO} frameworks in the mixOmics suite respectively use a sparse-PLS \citep{sparsepls} and PLS-based approach to identify those associations between features of two or more 'omics data sets that are the most informative to discriminate between phenotypes \citep{diablo,MINT}. 
Analogously, the \textit{microbe–metabolite vectors} (\textbf{mmvec}) algorithm intends to identify and estimate associations between microbes and metabolites using a neural network approach that explicitly addresses the compositional nature of the microbiome data \citep{mmvec}. 
However, these multivariate frameworks are data-driven and thus do not consider pre-existing knowledge structures, meaning that the most discriminative associations could be biologically meaningless.

\newpage

\section{Methods}

\subsection{Motivation}
The anansi framework relies on a \textit{knowledge-based binary adjacency matrix} to only assess associations between pairs of features that are known to interact in some fashion. 
The knowledge-based binary adjacency matrix is to used to mask associations that are not previously documented so that they can be skipped entirely for the purpose of downstream analysis, including visualisation, interpretation and indeed even hypothesis testing and subsequent multiple testing corrections. This application is key to addressing the aforementioned challenges of multi-omics integration: 
\begin{itemize}
    \item It solves the challenge of interpretability because resulting analysis is easier to interpret due to the structure imposed by the knowledge database. All remaining feature pairs will be structured and contextualized by their corresponding metabolic pathway in the knowledge database. 
    \item It solves the challenge of statistical power because power will be improved by avoiding unnecessary hypothesis testing in feature pairs that would be impossible to interpret due to the lack of a corresponding metabolic pathway in the knowledge database. By skipping non-canonical interactions in this manner, we preserve statistical power when applying FDR. 
\end{itemize}
Next, we will demonstrate how an all-vs-all association approach can be enhanced by introducing a knowledge-based binary adjacency matrix. 

\subsection{All-vs-all associations}
First, let us review an all-vs-all association analysis. \\
Suppose we have two 'omics data sets, \textbf{Y} and \textbf{X}, with $1...M_Y$ and $1...M_X$ features, respectively, both with $1...N$ samples. Each column within these data sets would contain the measured abundance of an 'omics feature, for which we represent the j-th feature of data set \textbf{Y} by

\begin{equation}
\textbf{f}_j^{(Y)} = [Y_{1j}, ..., Y_{Nj}]
\end{equation}
and analogously for \textbf{X}. 
\newline
Thus, we can view the data set \textbf{Y} as
\newline

\begin{equation}
\mathbf{Y} =
    \begin{blockarray}{ccccc}
         & \textbf{f}_{1}^{(Y)} & \textbf{f}_{2}^{(Y)} & \cdots & \textbf{f}_{M_Y}^{(Y)} \\
      \begin{block}{c[cccc]}
        & Y_{11} & Y_{12} & \cdots & Y_{1M_Y} \\
        & Y_{21} & Y_{22} & \cdots & Y_{2M_Y} \\
        & \vdots & \vdots & \ddots & \vdots \\
        & Y_{N1} & Y_{N2} & \cdots & Y_{NM_Y} \\
      \end{block}
    \end{blockarray}
\end{equation}
and analogously for \textbf{X}.
\newline
Here, columns represent different features and rows represent different samples or measurements. 
Then, the all-vs-all association matrix for these two data sets can be calculated as the association, $\rho$, between a column in \textbf{Y} and another column in \textbf{X}:
\newline

\begin{equation}
\rho(\mathbf{Y}, \mathbf{X}) :=
    \begin{blockarray}{ccccc}
      \begin{block}{c[cccc]}
        & \rho(\textbf{f}_1^{(Y)}, \textbf{f}_1^{(X)}) & \rho(\textbf{f}_1^{(Y)}, \textbf{f}_2^{(X)}) & \cdots & \rho(\textbf{f}_1^{(Y)}, \textbf{f}_{M_X}^{(X)}) \\ 
        & \rho(\textbf{f}_2^{(Y)}, \textbf{f}_1^{(X)}) & \rho(\textbf{f}_2^{(Y)}, \textbf{f}_2^{(X)}) & \cdots & \rho(\textbf{f}_2^{(Y)}, \textbf{f}_{M_X}^{(X)}) \\ 
        & \vdots & \vdots & \ddots & \vdots \\
        & \rho(\textbf{f}_{M_Y}^{(Y)}, \textbf{f}_1^{(X)}) & \rho(\textbf{f}_{M_Y}^{(Y)}, \textbf{f}_2^{(X)}) & \cdots & \rho(\textbf{f}_{M_Y}^{(Y)}, \textbf{f}_{M_X}^{(X)}) \\ 
      \end{block}
    \end{blockarray}
\end{equation}
\newline
Notice that the rows of our all-vs-all association matrix $\rho(\mathbf{Y}, \mathbf{X})$ correspond to the columns (features) in data set \textbf{Y} the columns of our association matrix $\rho(\mathbf{Y}, \mathbf{X})$ correspond to the columns from data set \textbf{X}. 
Further, notice that the resulting all-vs-all association matrix can be easily converted to a heatmap by depicting the resulting association coefficients a colour gradient. 

\newpage

\subsection{Binary adjacency matrix}

A binary adjacency matrix is a matrix of which the elements indicate which pairs of rows and columns are linked, or adjacent, to each other. It can be generated from an adjacency list, which in turn can be generated from a network graph such as those used to formulate a metabolic pathway network. Pairs of features between data sets that do interact, for instance based on whether they are connected in a \textit{knowledge database} like a metabolic pathway, will return a value of 1, whereas pairs of features that are not connected in such a manner will be depicted as 0. 
Put symbolically, a binary adjacency matrix \textbf{A} associates two sets $\textbf{u}=\{1, ..., U_{max}\}$ and $\textbf{t}=\{1, ..., T_{max}\}$, where

\begin{equation}
    A_{ij} = 
\begin{cases}
    1, & \text{if } u_i \text{ associates with } t_j \\
    0,              & \text{otherwise}
\end{cases}
\end{equation}

If we let each element in \textbf{u} represent one of the $M_Y$ features in \textbf{Y}, and each element in \textbf{t} represent one of the $M_X$ features in \textbf{X}, then 
\textbf{A} represents a binary adjacency matrix describing the relationship \textit{between} two data sets \textbf{Y} and \textbf{X}. 

For example, suppose data set \textbf{Y} contains metabolites from the KEGG database, whereas data set \textbf{X} contains molecular functions (KEGG orthologues). An adjacency list then contains information of which metabolites from \textbf{Y} are known to interact (i.e. are synthesised, catabolised, are cofactors of) with a function in \textbf{X}. This list, retrievable from publicly available databases, can be mapped into to create the binary adjacency matrix, as shown below:

\begin{equation}
\begin{tabular}{c c c}

    \begin{minipage}{.30 \linewidth}
        Adjacency list \\
        \\
        \begin{tabular}{l | l}
        \textbf{u} &  \textbf{t}\\
                    \hline
        $u_{1}$ & $\rightarrow$ $t_{1}$,$t_{2}$,$t_{3}$\\
        $u_{2}$ & $\rightarrow$ $t_{1}$,$t_{3}$,$t_{4}$\\
        $u_{3}$ & $\rightarrow$ $t_{1}$,$t_{4}$\\
        $u_{4}$ & $\rightarrow$ $t_{2}$,$t_{5}$\\
        $u_{5}$ & $\rightarrow$ $t_{1}$,$t_{3}$,$t_{4}$,$t_{5}$\\
        \end{tabular}\\ 
    \end{minipage} &

 \begin{minipage}{.1 \linewidth}
  $\rightarrow$ 
  
      \end{minipage}

    \begin{minipage}{.30\linewidth}
        Binary adjacency matrix\\
        \\
        \begin{tabular}{c |  c c c c c}
               & $t_{1}$ & $t_{2}$ & $t_{3}$ & $t_{4}$ & $t_{5}$  \\
            \hline
            $u_{1}$ & 1 & 1 & 1 & 0 & 0\\ 
            $u_{2}$ & 1 & 0 & 1 & 1 & 0\\ 
            $u_{3}$ & 1 & 0 & 0 & 1 & 0\\ 
            $u_{4}$ & 0 & 1 & 0 & 0 & 1\\ 
            $u_{5}$ & 1 & 0 & 1 & 1 & 1\\ 
        \end{tabular}
    \end{minipage}


\end{tabular}\\
\\
\end{equation}

\subsection{Masked association matrix}

Notice that since all elements from data set \textbf{Y} are listed as row names and all elements from data set \textbf{X} are listed as columns, the binary adjacency matrix will have the same dimensions as the association matrix. 
Because the dimensions are the same, the association matrix can be multiplied against the adjacency matrix to ``mask'' any associations that are not documented by the knowledge database. The masked association matrix, \textbf{R}, can thus be derived as follows:

\begin{equation} \label{eq:6}
\textbf{R} = \rho(\textbf{Y}, \textbf{X}) * \textbf{A}
\end{equation}

where
\begin{equation}
    R_{ij} =
    \begin{cases}
    \rho(\textbf{f}^{(Y)}_i, \textbf{f}^{(X)}_j), & \text{if } u_i \text{ associates with } t_j \\
    0,              & \text{otherwise}
\end{cases}
\end{equation}

The masked association matrix \textbf{R} should be seen as the primary output of the anansi package and can serve as the basis for follow-up analyses such as differential association analysis, network analysis, and functional enrichment analysis. 

\newpage


\subsection{Implementation}
The anansi framework was written in the R language. The general workflow of this package can be conceptualized as a three-step process: 
\begin{itemize}
    \item Generate a binary adjacency matrix with the appropriate dimensions and features by cross-referencing the feature table(s) with the knowledge database.
    
    \item Compute the masked association matrix \textbf{R} by multiplying $\rho(\textbf{Y}, \textbf{X})$ with $\textbf{A}$ (See Equation \ref{eq:6}). 

    \item Perform follow-up analyses on the resulting masked association matrix \textbf{R}, such as differential association analysis.   
\end{itemize}
An overview of the internal architecture can be seen in figure 1. Statistics are computed using the base R stats package and the lme4 package for linear mixed effects models \citep{R,lme4}. ``getter'' functions are available to parse output files into the wide format as well as the long format, designed to be compatible with the ggplot2 plotting software \citep{tidydata,ggplot2}. Anansi is parallelizable by using the futures framework \citep{futuresR}.

\begin{figure}[hbtp]
\includegraphics[width=\textwidth]{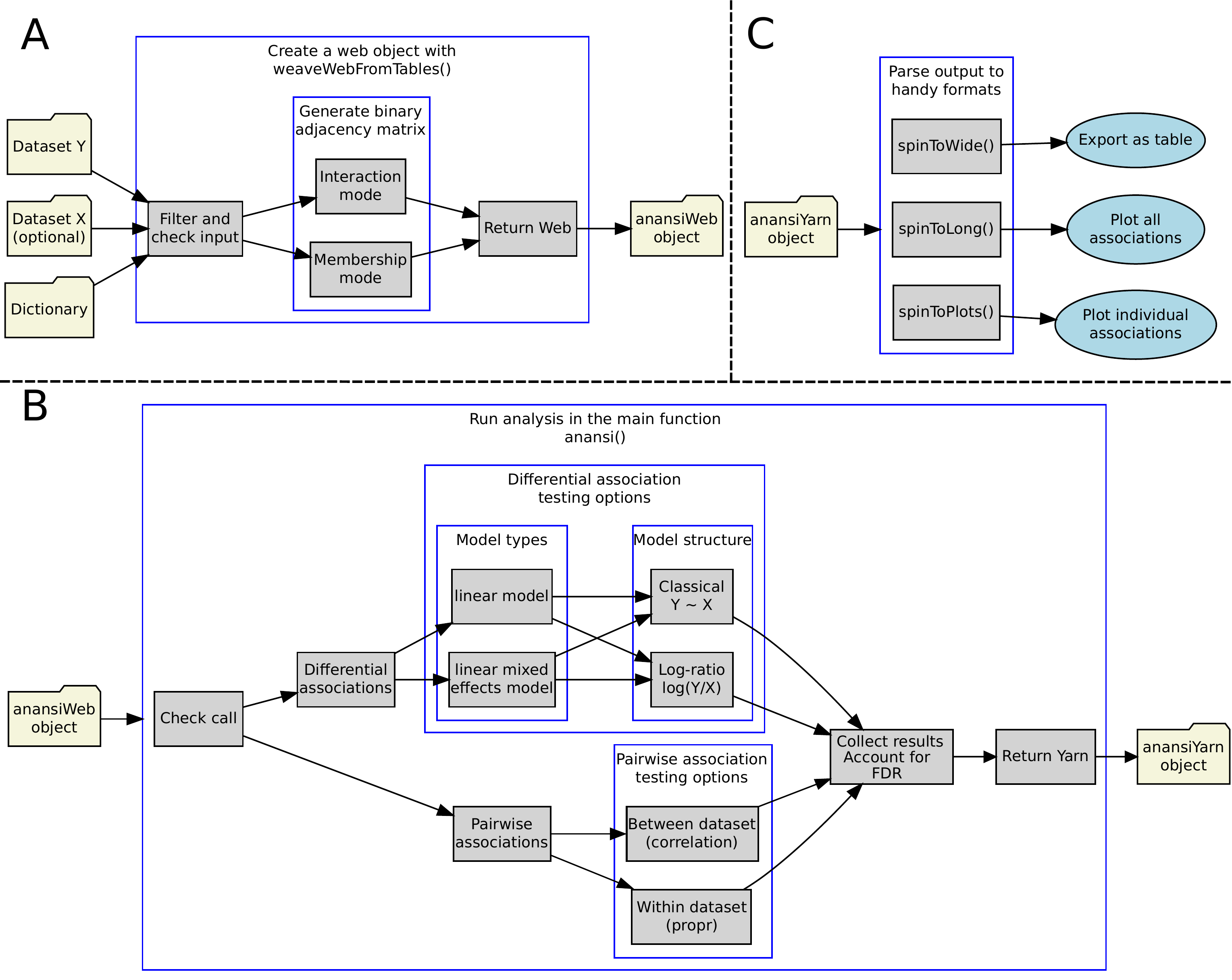}    
\centering
\caption{Diagram  of the anansi workflow. The anansi workflow relies on three steps, each with their own function and custom S4 object classes. In step \textbf{A}, input 'omics data sets are collected and compared to a knowledge-based adjacency list, referred to here as a dictionary. The adjacency lists can be based on known interaction, such as in a metabolic network, or based on membership of a shared overarching category such as a gene pathway. Features that are not part of at least one pair are omitted here. An anansiWeb S4 object is returned, which is the input for the main anansi function. In \textbf{B}, the masked association matrix \textbf{R} is computed. Optionally, differential associations are assessed. All results are collected and adjusted p-values are computed, after which an S4 anansiYarn object is returned, which can be parsed into different formats by the functions in \textbf{C}. Results can be parsed to a long format directly compatible with the ggplot2 plotting software as well as to a publication-ready wide format table.}

\end{figure}

\section{Results}
\subsection{Differential association testing}

In order to assess differences in associations based on one or more variables (such as phenotype or treatment), we make use of the emergent and disjointed association paradigm introduced in the context of proportionality \citep{diff_prop,quinn2017propr,part_corr} and apply it outside of the simplex. Briefly, disjointed associations refer to the scenario where the \textit{slope} of an association is dependent on a variable. On the other hand, emergent associations refer to the scenario where the \textit{strength} of the scenario is dependent on a variable. See figure 2 for an illustrated hypothetical example. Anansi supports arbitrarily complex linear models as well as longitudinal models using formula syntax from the base R stats package and the lme4 package for linear mixed effects models, respectively \citep{R,lme4}. 

\begin{figure}[hbtp]
\includegraphics[width=\textwidth]{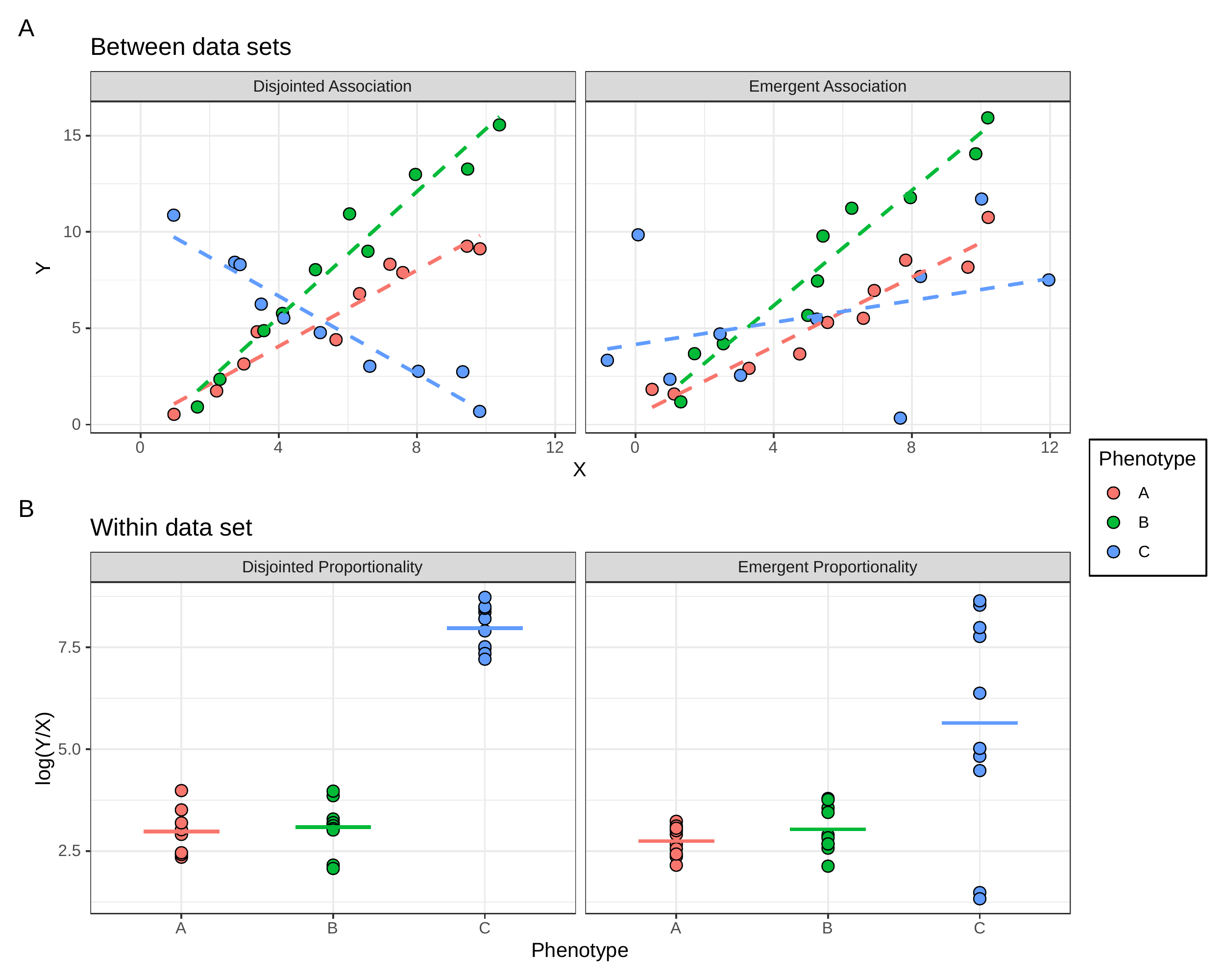}    
\centering
\caption{An example of differential associations between hypothetical features Y and X. In all cases, phenotype C illustrates the differential association compared to phenotypes A \& B. Disjointed associations describe the scenario where there is a detectable association in all cases, but the quality of the association differs. Emergent associations describe the case where an association can be detected in one case but not in another. In scenario \textbf{A}, the features Y and X are from different datasets and differential associations can be assessed using a classical linear models: $lm(Y \sim X \underline{\times Phenotype})$ and $lm(residuals(lm(Y \sim X) \sim \underline{Phenotype})$ for disjointed and emergent associations, respectively. In scenario \textbf{B}, the features are from the same compositional dataset. Differential proportionality can be assessed using applying similar models on log-ratios: $lm(log(\frac{Y}{X}) \sim \underline{Phenotype})$ and $lm(residuals(lm(log(\frac{Y}{X}) \sim 1) \sim \underline{Phenotype})$ for disjointed and emergent proportionality, respectively. In all cases, the $R^2$ and p-value for the underlined part of the equation is considered to estimate differential associations. }

\end{figure}
\newpage



\subsection{FMT Ageing}
An early version of the anansi framework was used in a recent publication, assessing the associations between hippocampal metabolites and microbial functions \citep[Extended Data Fig.~7]{FMT_aging}. Briefly, the aim of the study was to investigate whether faecal microbiota transplantation from young donor mice could restore symptoms of ageing in aged recipient mice. As part of the study, the functional metagenome was inferred from 16S rRNA sequencing data and compared to hippocampal metabolite levels. Hippocampal metabolites were first linked to microbial functions that either produce or metabolise these metabolites. Then, association strength was assessed for each of these pairs, both per treatment group and between treatment groups. Strikingly, the slope of the associations between feature pairs, such as lactate vs lactate dehydrogenase, was completely inverted, implying that the relation between these features is dependent on the treatment received (Figure 3). The specific nature of these results enables researchers to formulate follow-up hypotheses. The anansi package contains curated snippets of this dataset for tutorial purposes. Full analysis is available online: \url{https://github.com/thomazbastiaanssen/anansi/} 

\begin{figure}[hbtp]
\includegraphics[width=\textwidth]{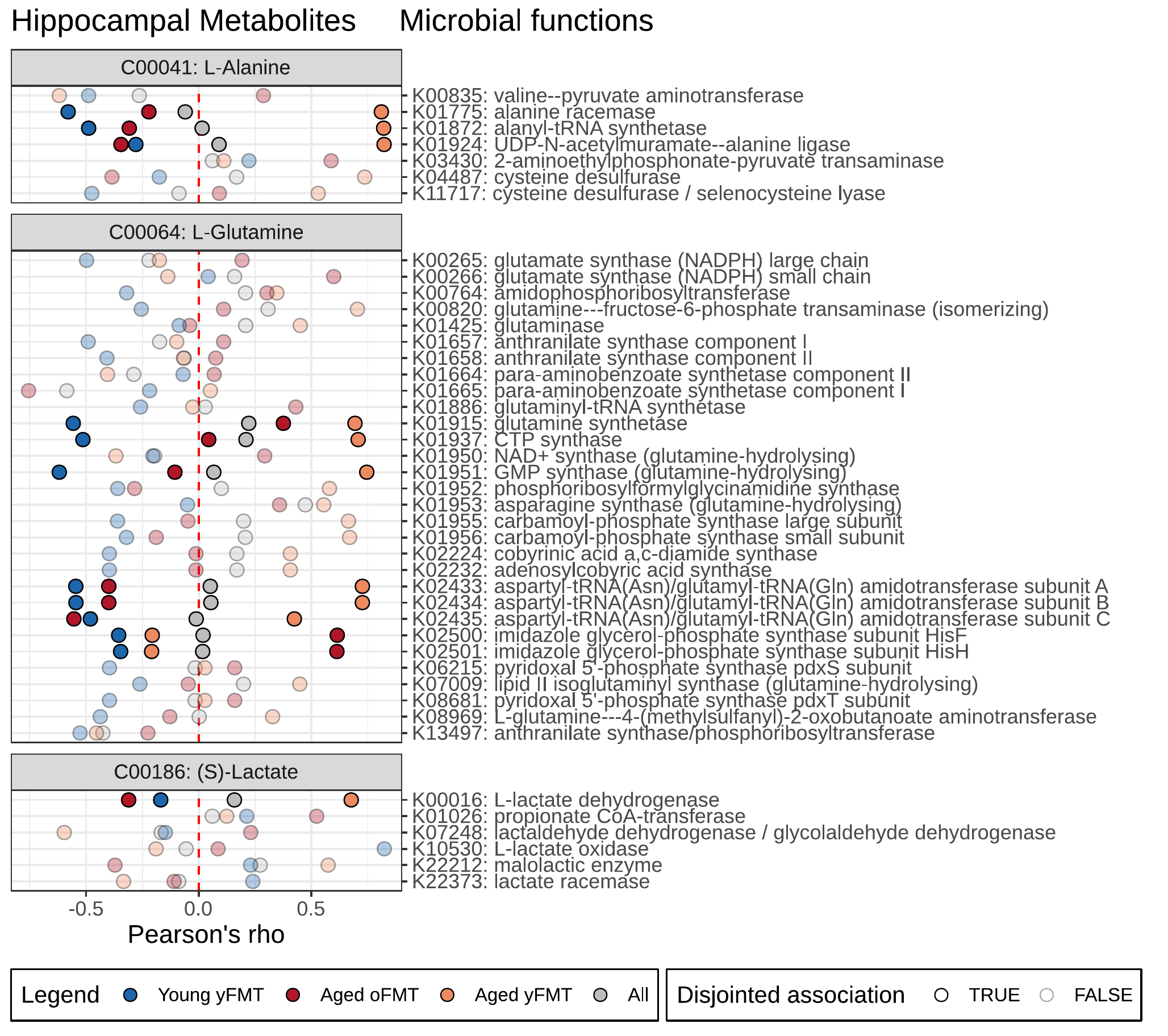}    
\centering
\caption{Figure showing the associations between hippocampal metabolite levels and related microbial functions. X-axis shows Pearson correlation coefficients for the the metabolite-function pairs. The red vertical dashed lines depict a Pearson correlation coefficient of 0. Colours depict the treatment groups, whereas grey points represent the correlation coefficients after pooling all three groups. Opaque points with black borders display significantly disjointed associations. Non-opaque points display the correlation coefficient for the associations where the full model fitted sufficiently well after FDR, but where no disjointed associations were detected.}

\end{figure}

\newpage

\section{Discussion}

\subsection{Towards Interpretability}
We have argued that there are two main challenges when using multi-omics integration analysis to formulate testable hypotheses, namely 1) interpretation of results and 2) the preservation of statistical power. The anansi framework addresses both of these challenges by constraining the all-vs-all association hypothesis space, only considering pairwise associations that are \textit{a priori} known to occur. This constraint guarantees that all resulting associations occur in the knowledge database and that no statistical power is wasted by unnecessarily extending FDR adjusting to undocumented -and thus likely uninterpretable- associations. 

\subsection{Limitations}
There are a few limitations to consider when using anansi. First, functional metagenomics data, such as the output from PICRUSt2 \citep{picrust2} and HUMAnN3 \citep{humann3} is nested in the sense that the abundance of each function is directly dependent on the abundance on the respective taxa that contain those genes, which may lead to violations of independence between features for the purpose of controlling the false discovery rate (FDR) and lead to spurious associations. A move towards metatranscriptomics and/or metaproteomics rather than metagenomics would alleviate the by-taxon dependence between functions. 

Second, the accuracy of anansi is highly dependent on the accuracy of the databases used to generate an adjacency matrix. Feature pairs that interact in reality but are not catalogued as such will not be assessed. Interestingly, in a recent study attempting to link the levels of serum metabolites to a variety of factors including the mcirobiome, many unknown metabolites and xenobiotics were linked to the microbiome \citep{metabolomics_map}. It stands to reason that, especially when it comes to interkingdom communication, many associations have simply not been mapped. That said, it is unlikely that associations between feature pairs that in reality do interact, yet have not been catalogued as such, will lead to fruitful hypotheses even if they were assessed in the context of an integratomics analysis. 

Third, anansi currently only supports a binary adjacency matrix, but in biology, interaction is often on a spectrum. For instance, different ligands bind to their respective receptors at different efficiencies. Future implementations using a knowledge-based adjacency matrix may expand on this principle by allowing for continuous interaction scores. 

\subsection{Conclusions}
While the anansi framework was designed with microbiome and metabolomics data in mind, it could feasibly be applied any field where interactions between two large data sets where only some features meaningfully interact need to be assessed. Example applications include phage-bacterium, immune-metabolite or receptor-ligand interaction analysis. \\
As 'omics data sets increase in number and complexity, there is a dire need for tools and approaches to process and parse this data in such a way that meaningful and testable hypotheses can be formulated. The microbiome field is in need of methods to investigate causality \citep{causality_bastiaanssen, causality_cryan} and we view anansi as one of many approaches to move towards this goal. 

\section{Acknowledgements}
APC Microbiome Ireland is a research centre funded by Science Foundation Ireland (SFI), through the Irish Governments’ national development plan (grant no. 12/RC/2273\_P2). \\
We are grateful for the helpful comments and encouragement of Aonghus Lavelle, Benjamin Valderrama, Frank Snijders, Ionas Erb and Sarah-Jane Leigh. The anansi hexagon sticker was designed by Johanna Snijders (nightillu). 

\section{Declarations}

TFSB and TPQ declare no competing interests. JFC has been an invited speaker at conferences organized by Mead Johnson, Ordesa, and Yakult, and has received research funding from Reckitt, Nutricia, Dupont/IFF, and Nestle. This did not influence this manuscript in any way. 

\section{Code Availability}
Anansi is open source and freely available under the GPL-3 licence. The code implementing the anansi algorithm as well as a tutorial demonstrating the analysis performed for the FMT Ageing manuscript can be found on GitHub: \url{https://github.com/thomazbastiaanssen/anansi}

\bibliography{anansi}

\begin{thebibliography}{33}
\providecommand{\natexlab}[1]{#1}
\providecommand{\url}[1]{\texttt{#1}}
\expandafter\ifx\csname urlstyle\endcsname\relax
  \providecommand{\doi}[1]{doi: #1}\else
  \providecommand{\doi}{doi: \begingroup \urlstyle{rm}\Url}\fi

\bibitem[Bar et~al.(2020)Bar, Korem, Weissbrod, Zeevi, Rothschild, Leviatan,
  Kosower, Lotan-Pompan, Weinberger, Le~Roy, et~al.]{metabolomics_map}
N.~Bar, T.~Korem, O.~Weissbrod, D.~Zeevi, D.~Rothschild, S.~Leviatan,
  N.~Kosower, M.~Lotan-Pompan, A.~Weinberger, C.~I. Le~Roy, et~al.
\newblock A reference map of potential determinants for the human serum
  metabolome.
\newblock \emph{Nature}, 588\penalty0 (7836):\penalty0 135--140, 2020.

\bibitem[Bastiaanssen and Cryan(2021)]{causality_bastiaanssen}
T.~F. Bastiaanssen and J.~F. Cryan.
\newblock The microbiota-gut-brain axis in mental health and medication
  response: parsing directionality and causality.
\newblock \emph{International Journal of Neuropsychopharmacology}, 24\penalty0
  (3):\penalty0 216--220, 2021.

\bibitem[Bates et~al.(2015)Bates, M{\"a}chler, Bolker, and Walker]{lme4}
D.~Bates, M.~M{\"a}chler, B.~Bolker, and S.~Walker.
\newblock Fitting linear mixed-effects models using {lme4}.
\newblock \emph{Journal of Statistical Software}, 67\penalty0 (1):\penalty0
  1--48, 2015.
\newblock \doi{10.18637/jss.v067.i01}.

\bibitem[Beghini et~al.(2021)Beghini, McIver, Blanco-Míguez, Dubois, Asnicar,
  Maharjan, Mailyan, Manghi, Scholz, Thomas, Valles-Colomer, Weingart, Zhang,
  Zolfo, Huttenhower, Franzosa, and Segata]{humann3}
F.~Beghini, L.~J. McIver, A.~Blanco-Míguez, L.~Dubois, F.~Asnicar,
  S.~Maharjan, A.~Mailyan, P.~Manghi, M.~Scholz, A.~M. Thomas,
  M.~Valles-Colomer, G.~Weingart, Y.~Zhang, M.~Zolfo, C.~Huttenhower, E.~A.
  Franzosa, and N.~Segata.
\newblock Integrating taxonomic, functional, and strain-level profiling of
  diverse microbial communities with biobakery 3.
\newblock \emph{eLife}, 10:\penalty0 e65088, may 2021.
\newblock ISSN 2050-084X.
\newblock \doi{10.7554/eLife.65088}.
\newblock URL \url{https://doi.org/10.7554/eLife.65088}.

\bibitem[Bengtsson(2021)]{futuresR}
H.~Bengtsson.
\newblock A unifying framework for parallel and distributed processing in r
  using futures.
\newblock \emph{The R Journal}, 13\penalty0 (2):\penalty0 208--227, 2021.
\newblock \doi{10.32614/RJ-2021-048}.
\newblock URL \url{https://doi.org/10.32614/RJ-2021-048}.

\bibitem[Benjamini and Hochberg(1995)]{BH}
Y.~Benjamini and Y.~Hochberg.
\newblock Controlling the false discovery rate: A practical and powerful
  approach to multiple testing.
\newblock \emph{Journal of the Royal Statistical Society. Series B
  (Methodological)}, 57\penalty0 (1):\penalty0 289--300, 1995.
\newblock ISSN 00359246.
\newblock URL \url{http://www.jstor.org/stable/2346101}.

\bibitem[Boehme et~al.(2021)Boehme, Guzzetta, Bastiaanssen, van~de Wouw,
  Moloney, Gual-Grau, Spichak, Olavarr{\'{\i}}a-Ram{\'{\i}}rez, Fitzgerald,
  Morillas, Ritz, Jaggar, Cowan, Crispie, Donoso, Halitzki, Neto, Sichetti,
  Golubeva, Fitzgerald, Claesson, Cotter, O'Leary, Dinan, and Cryan]{FMT_aging}
M.~Boehme, K.~E. Guzzetta, T.~F.~S. Bastiaanssen, M.~van~de Wouw, G.~M.
  Moloney, A.~Gual-Grau, S.~Spichak, L.~Olavarr{\'{\i}}a-Ram{\'{\i}}rez,
  P.~Fitzgerald, E.~Morillas, N.~L. Ritz, M.~Jaggar, C.~S.~M. Cowan,
  F.~Crispie, F.~Donoso, E.~Halitzki, M.~C. Neto, M.~Sichetti, A.~V. Golubeva,
  R.~S. Fitzgerald, M.~J. Claesson, P.~D. Cotter, O.~F. O'Leary, T.~G. Dinan,
  and J.~F. Cryan.
\newblock Microbiota from young mice counteracts selective age-associated
  behavioral deficits.
\newblock \emph{Nature Aging}, 1\penalty0 (8):\penalty0 666--676, Aug. 2021.
\newblock \doi{10.1038/s43587-021-00093-9}.
\newblock URL \url{https://doi.org/10.1038/s43587-021-00093-9}.

\bibitem[Caspi et~al.(2014)Caspi, Altman, Billington, Dreher, Foerster,
  Fulcher, Holland, Keseler, Kothari, Kubo, et~al.]{metacyc}
R.~Caspi, T.~Altman, R.~Billington, K.~Dreher, H.~Foerster, C.~A. Fulcher,
  T.~A. Holland, I.~M. Keseler, A.~Kothari, A.~Kubo, et~al.
\newblock The metacyc database of metabolic pathways and enzymes and the biocyc
  collection of pathway/genome databases.
\newblock \emph{Nucleic acids research}, 42\penalty0 (D1):\penalty0 D459--D471,
  2014.

\bibitem[Cryan and Mazmanian(2022)]{causality_cryan}
J.~F. Cryan and S.~K. Mazmanian.
\newblock Microbiota--brain axis: Context and causality.
\newblock \emph{Science}, 376\penalty0 (6596):\penalty0 938--939, 2022.

\bibitem[Douglas et~al.(2020)Douglas, Maffei, Zaneveld, Yurgel, Brown, Taylor,
  Huttenhower, and Langille]{picrust2}
G.~M. Douglas, V.~J. Maffei, J.~R. Zaneveld, S.~N. Yurgel, J.~R. Brown, C.~M.
  Taylor, C.~Huttenhower, and M.~G. Langille.
\newblock Picrust2 for prediction of metagenome functions.
\newblock \emph{Nature biotechnology}, 38\penalty0 (6):\penalty0 685--688,
  2020.

\bibitem[Drula et~al.(2022)Drula, Garron, Dogan, Lombard, Henrissat, and
  Terrapon]{cazy2022}
E.~Drula, M.-L. Garron, S.~Dogan, V.~Lombard, B.~Henrissat, and N.~Terrapon.
\newblock The carbohydrate-active enzyme database: functions and literature.
\newblock \emph{Nucleic acids research}, 50\penalty0 (D1):\penalty0 D571--D577,
  2022.

\bibitem[Erb(2020)]{part_corr}
I.~Erb.
\newblock Partial correlations in compositional data analysis.
\newblock \emph{Applied Computing and Geosciences}, 6:\penalty0 100026, 2020.
\newblock ISSN 2590-1974.
\newblock \doi{https://doi.org/10.1016/j.acags.2020.100026}.
\newblock URL
  \url{https://www.sciencedirect.com/science/article/pii/S2590197420300082}.

\bibitem[Erb et~al.(2017)Erb, Quinn, Lovell, and Notredame]{diff_prop}
I.~Erb, T.~Quinn, D.~Lovell, and C.~Notredame.
\newblock Differential proportionality - a normalization-free approach to
  differential gene expression.
\newblock \emph{bioRxiv}, May 2017.
\newblock \doi{10.1101/134536}.
\newblock URL \url{https://doi.org/10.1101/134536}.

\bibitem[Friedman and Alm(2012)]{SPARCC}
J.~Friedman and E.~J. Alm.
\newblock Inferring correlation networks from genomic survey data.
\newblock \emph{PLoS computational biology}, 8\penalty0 (9):\penalty0 e1002687,
  2012.

\bibitem[Ghazi et~al.(2021)Ghazi, Sucipto, Rahnavard, Franzosa, McIver,
  Lloyd-Price, Schwager, Weingart, Moon, Morgan, Waldron, and
  Huttenhower]{halla}
A.~R. Ghazi, K.~Sucipto, G.~Rahnavard, E.~A. Franzosa, L.~J. McIver,
  J.~Lloyd-Price, E.~Schwager, G.~Weingart, Y.~S. Moon, X.~C. Morgan,
  L.~Waldron, and C.~Huttenhower.
\newblock High-sensitivity pattern discovery in large, paired multi-omic
  datasets.
\newblock \emph{bioRxiv}, 2021.
\newblock \doi{10.1101/2021.11.11.468183}.
\newblock URL
  \url{https://www.biorxiv.org/content/early/2021/11/13/2021.11.11.468183}.

\bibitem[Gloor et~al.(2017)Gloor, Macklaim, Pawlowsky-Glahn, and
  Egozcue]{gloor_frontiers}
G.~B. Gloor, J.~M. Macklaim, V.~Pawlowsky-Glahn, and J.~J. Egozcue.
\newblock Microbiome datasets are compositional: And this is not optional.
\newblock \emph{Front Microbiol}, 8:\penalty0 2224, 2017.
\newblock ISSN 1664-302X (Print) 1664-302x.
\newblock \doi{10.3389/fmicb.2017.02224}.

\bibitem[Huerta-Cepas et~al.(2019)Huerta-Cepas, Szklarczyk, Heller,
  Hern{\'a}ndez-Plaza, Forslund, Cook, Mende, Letunic, Rattei, Jensen,
  et~al.]{eggnog}
J.~Huerta-Cepas, D.~Szklarczyk, D.~Heller, A.~Hern{\'a}ndez-Plaza, S.~K.
  Forslund, H.~Cook, D.~R. Mende, I.~Letunic, T.~Rattei, L.~J. Jensen, et~al.
\newblock eggnog 5.0: a hierarchical, functionally and phylogenetically
  annotated orthology resource based on 5090 organisms and 2502 viruses.
\newblock \emph{Nucleic acids research}, 47\penalty0 (D1):\penalty0 D309--D314,
  2019.

\bibitem[Kanehisa and Goto(2000)]{kegg}
M.~Kanehisa and S.~Goto.
\newblock Kegg: kyoto encyclopedia of genes and genomes.
\newblock \emph{Nucleic acids research}, 28\penalty0 (1):\penalty0 27--30,
  2000.

\bibitem[Kenney and Keeping(1962)]{kenney1962linear}
J.~F. Kenney and E.~Keeping.
\newblock Linear regression and correlation.
\newblock \emph{Mathematics of statistics}, 1:\penalty0 252--285, 1962.

\bibitem[Kurtz et~al.(2015)Kurtz, M{\"u}ller, Miraldi, Littman, Blaser, and
  Bonneau]{SPieCeasi}
Z.~D. Kurtz, C.~L. M{\"u}ller, E.~R. Miraldi, D.~R. Littman, M.~J. Blaser, and
  R.~A. Bonneau.
\newblock Sparse and compositionally robust inference of microbial ecological
  networks.
\newblock \emph{PLoS computational biology}, 11\penalty0 (5):\penalty0
  e1004226, 2015.

\bibitem[L{\^e}~Cao et~al.(2011)L{\^e}~Cao, Boitard, and Besse]{sparsepls}
K.-A. L{\^e}~Cao, S.~Boitard, and P.~Besse.
\newblock Sparse pls discriminant analysis: biologically relevant feature
  selection and graphical displays for multiclass problems.
\newblock \emph{BMC bioinformatics}, 12\penalty0 (1):\penalty0 1--17, 2011.

\bibitem[Lovell et~al.(2015)Lovell, Pawlowsky-Glahn, Egozcue, Marguerat, and
  B{\"a}hler]{lovell2015proportionality}
D.~Lovell, V.~Pawlowsky-Glahn, J.~J. Egozcue, S.~Marguerat, and J.~B{\"a}hler.
\newblock Proportionality: a valid alternative to correlation for relative
  data.
\newblock \emph{PLoS computational biology}, 11\penalty0 (3):\penalty0
  e1004075, 2015.

\bibitem[Morton et~al.(2019)Morton, Aksenov, Nothias, Foulds, Quinn, Badri,
  Swenson, Van~Goethem, Northen, Vazquez-Baeza, et~al.]{mmvec}
J.~T. Morton, A.~A. Aksenov, L.~F. Nothias, J.~R. Foulds, R.~A. Quinn, M.~H.
  Badri, T.~L. Swenson, M.~W. Van~Goethem, T.~R. Northen, Y.~Vazquez-Baeza,
  et~al.
\newblock Learning representations of microbe--metabolite interactions.
\newblock \emph{Nature methods}, 16\penalty0 (12):\penalty0 1306--1314, 2019.

\bibitem[Quinn et~al.(2017)Quinn, Richardson, Lovell, and
  Crowley]{quinn2017propr}
T.~P. Quinn, M.~F. Richardson, D.~Lovell, and T.~M. Crowley.
\newblock propr: an r-package for identifying proportionally abundant features
  using compositional data analysis.
\newblock \emph{Scientific reports}, 7\penalty0 (1):\penalty0 1--9, 2017.

\bibitem[{R Core Team}(2022)]{R}
{R Core Team}.
\newblock \emph{R: A Language and Environment for Statistical Computing}.
\newblock R Foundation for Statistical Computing, Vienna, Austria, 2022.
\newblock URL \url{https://www.R-project.org/}.

\bibitem[Rohart et~al.(2017)Rohart, Eslami, Matigian, Bougeard, and
  Le~Cao]{MINT}
F.~Rohart, A.~Eslami, N.~Matigian, S.~Bougeard, and K.-A. Le~Cao.
\newblock Mint: a multivariate integrative method to identify reproducible
  molecular signatures across independent experiments and platforms.
\newblock \emph{BMC bioinformatics}, 18\penalty0 (1):\penalty0 1--13, 2017.

\bibitem[Singh et~al.(2019)Singh, Shannon, Gautier, Rohart, Vacher, Tebbutt,
  and Lê~Cao]{diablo}
A.~Singh, C.~P. Shannon, B.~Gautier, F.~Rohart, M.~Vacher, S.~J. Tebbutt, and
  K.-A. Lê~Cao.
\newblock {DIABLO: an integrative approach for identifying key molecular
  drivers from multi-omics assays}.
\newblock \emph{Bioinformatics}, 35\penalty0 (17):\penalty0 3055--3062, 01
  2019.
\newblock ISSN 1367-4803.
\newblock \doi{10.1093/bioinformatics/bty1054}.
\newblock URL \url{https://doi.org/10.1093/bioinformatics/bty1054}.

\bibitem[Storey and Tibshirani(2003)]{qvalue}
J.~D. Storey and R.~Tibshirani.
\newblock Statistical significance for genomewide studies.
\newblock \emph{Proceedings of the National Academy of Sciences}, 100\penalty0
  (16):\penalty0 9440--9445, 2003.
\newblock \doi{10.1073/pnas.1530509100}.
\newblock URL \url{https://www.pnas.org/doi/abs/10.1073/pnas.1530509100}.

\bibitem[Subramanian et~al.(2020)Subramanian, Verma, Kumar, Jere, and
  Anamika]{multi_omics}
I.~Subramanian, S.~Verma, S.~Kumar, A.~Jere, and K.~Anamika.
\newblock Multi-omics data integration, interpretation, and its application.
\newblock \emph{Bioinformatics and Biology Insights}, 14:\penalty0
  1177932219899051, 2020.
\newblock \doi{10.1177/1177932219899051}.
\newblock URL \url{https://doi.org/10.1177/1177932219899051}.
\newblock PMID: 32076369.

\bibitem[Wickham(2014)]{tidydata}
H.~Wickham.
\newblock Tidy data.
\newblock \emph{Journal of Statistical Software}, 59\penalty0 (10):\penalty0
  1–23, 2014.
\newblock \doi{10.18637/jss.v059.i10}.
\newblock URL
  \url{https://www.jstatsoft.org/index.php/jss/article/view/v059i10}.

\bibitem[Wickham(2016)]{ggplot2}
H.~Wickham.
\newblock \emph{ggplot2: Elegant Graphics for Data Analysis}.
\newblock Springer-Verlag New York, 2016.
\newblock ISBN 978-3-319-24277-4.
\newblock URL \url{https://ggplot2.tidyverse.org}.

\bibitem[Wishart et~al.(2007)Wishart, Tzur, Knox, Eisner, Guo, Young, Cheng,
  Jewell, Arndt, Sawhney, et~al.]{hmdb}
D.~S. Wishart, D.~Tzur, C.~Knox, R.~Eisner, A.~C. Guo, N.~Young, D.~Cheng,
  K.~Jewell, D.~Arndt, S.~Sawhney, et~al.
\newblock Hmdb: the human metabolome database.
\newblock \emph{Nucleic acids research}, 35\penalty0 (suppl\_1):\penalty0
  D521--D526, 2007.

\bibitem[Yanai and Lercher(2019)]{night_science}
I.~Yanai and M.~Lercher.
\newblock Night science.
\newblock \emph{Genome Biology}, 20\penalty0 (1):\penalty0 1--3, 2019.

\end{thebibliography}

\end{document}